\documentclass[aps,prl,reprint,groupedaddress]{revtex4-2}

\usepackage{graphicx}  
\usepackage{dcolumn}   
\usepackage{bm}        
\usepackage{amssymb}   
\usepackage{amsmath}
\usepackage{amsfonts}
\usepackage{amssymb}

\usepackage{xcolor}
\PassOptionsToPackage{normalem}{ulem}
\usepackage{ulem}

\providecolor{added}{rgb}{0,0,1}
\providecolor{deleted}{rgb}{1,0,0}

\newcommand{\bs}{\boldsymbol}

\newcommand{\s}{\sigma}

\newcommand{\spx}{\mathbf{x}}

\begin{document}

\title{Observation of transition radiation carrying orbital angular momentum}

\author{Y.~Takabayashi}
\email{takabayashi@saga-ls.jp}
\author{H.~Takeda}
\author{E.~Magome}
\affiliation{SAGA Light Source, 8-7 Yayoigaoka, Tosu, Saga 841-0005, Japan}
\author{K.~Sumitani}
\affiliation{Japan Synchrotron Radiation Research Institute (JASRI), 1-1-1 Kouto, Sayo-cho, Sayo-gun, Hyogo 679-5198, Japan}
\author{P.~O.~Kazinski}
\affiliation{Physics Faculty, Tomsk State University, Tomsk 634050, Russia}
\author{P.~S.~Korolev}
\affiliation{Physics Faculty, Tomsk State University, Tomsk 634050, Russia}
\author{O.~V.~Bogdanov}
\email{bov@tpu.ru}
\affiliation{Mathematics and Mathematical Physics Division, National Research Tomsk Polytechnic University, Tomsk 634050, Russia}
\author{T.~A.~Tukhfatullin}
\email{tta@tpu.ru}
\affiliation{Almaty branch of National Research Nuclear University MEPhI, Almaty 050040, Kazakhstan}

\date{\today}

\begin{abstract}
Twisted photons carrying orbital angular momentum, which have potential applications spanning diverse fields, have been extensively studied since the theoretical work of Allen \textit{et al}. in 1992.
Various methods for direct producing twisted photons have been explored, leveraging the rotational (spiral) motion of relativistic electrons in phenomena such as undulator radiation.
In the present study, transition radiation carrying orbital angular momentum is observed for the first time.
This radiation was generated by 220 MeV electrons incident on an Au-coated Si wafer.
The orbital angular momentum was measured by analyzing the diffraction patterns produced as the radiation passed through a triangular aperture and a double slit.
These results demonstrate that twisted photons can also be generated through the interaction of rectilinearly moving electrons with a solid target.
\end{abstract}

\maketitle
Since the seminal work of Allen \textit{et al}. in 1992 \cite{Allen_PRA1992}, twisted photons have been the subject of intensive theoretical and experimental studies \cite{rubinsztein2016roadmap,Padgett_OptExp2017}.
Twisted photons can carry orbital angular momentum (OAM), which, in principle, can assume arbitrary values with $\hbar l$ ($|l| \geq$ 1), unlike spin angular momentum (SAM) $\hbar s$, which is limited to $|s| = 1$,
where $\hbar$ is the Planck constant divided by 2$\pi$.
In addition, the phase of twisted photons exhibits a singularity at the center, resulting in zero intensity and a characteristic doughnut-shaped angular distribution.
These unique features have enabled numerous potential applications, including optical tweezers \cite{He_PRL1995,Grier_Nature2003}, compression of beams of charged particles by twisted photons \cite{willim2023proton,jirka2024effects}, super-resolution microscopy \cite{Hell_OptLett1994,Torok_OptExp2004}, material modification \cite{Toyoda_PRL2013}, twisted surface photoelectric effect \cite{kazinski2024surface}, photoexcitation of transitions with high multipolarity \cite{afanasev2013off,lu2023manipulation,kazinski2024excitation,kirschbaum2024photoexcitation}, and channel multiplexing in telecommunications \cite{Gibson_OptExp2004,Wang_NatPho2012,jiang2021electromagnetic,noor2022review,kazinski2024multiplexing}.

On the other hand, studies aimed at producing twisted photons have been conducted using relativistic electrons to cover a wider spectral range, particularly at higher photon energies, and to increase the intensity of radiation.
One approach involves employing undulators \cite{Sasaki_PRL2008,Bahrdt_PRL2013,Katoh_SciRep2017}.
In circularly polarized undulators, electrons follow a spiral trajectory; although the fundamental undulator radiation carries SAM, higher-order undulator radiation is known to carry OAM.
Another approach utilizes laser Compton scattering \cite{Jentschura_PRL2011,Taira_SciRep2017,Sakai_PRAB2015}.
Jentschura and Serbo \cite{Jentschura_PRL2011} predicted that twisted $\gamma$-rays will be produced from collisions between high-energy electrons and twisted photons.
By contrast, other authors \cite{Taira_SciRep2017,Sakai_PRAB2015,bogdanov2019semiclassical} predicted that twisted $\gamma$-rays will be produced via nonlinear Compton scattering involving high-intensity lasers.
Channeling radiation has also been theoretically predicted to carry OAM \cite{Abdrashitov_PLA2018,Epp_NIMB2018}.
When electrons (or positrons) are axially channeled in a single crystal, they spiral along the crystal axis, emitting $\gamma$-rays with OAM.
However, this prediction has not been experimentally confirmed.
In addition, twisted photons can be produced in an undulator if the electron beam is helically microbunched \cite{Hemsing_APL2012,Hemsing_NatPhys2013}.
In this scheme, such a bunch is shaped through interactions between the electrons in the undulator and a laser beam introduced into the undulator in the same direction as the electron beam.

\begin{figure*}[t]
\includegraphics[width=0.7\hsize]{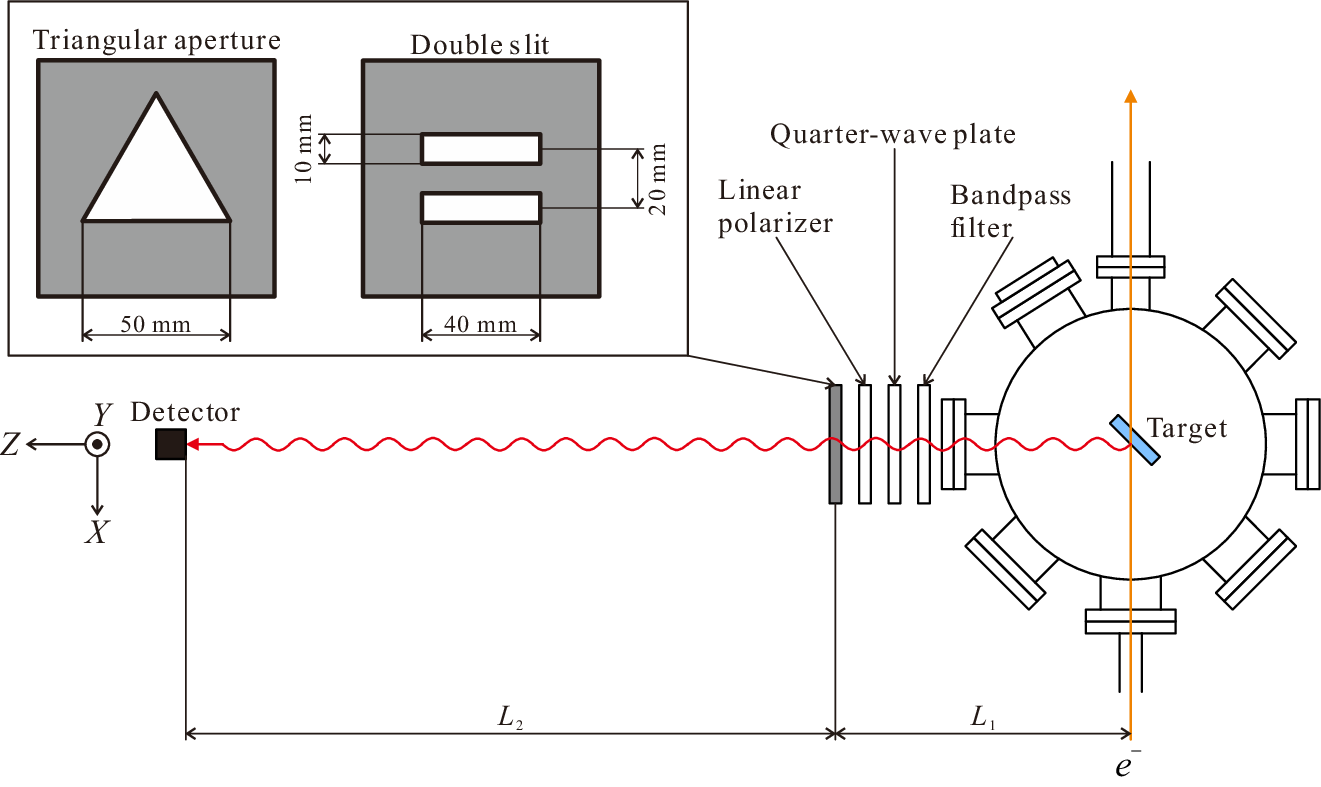}
\caption{A schematic of the experimental setup is shown. The inset provides detailed illustrations of the triangular aperture and the double-slit configurations.
For the triangular aperture experiment, $L_1=500$ mm and $L_2=1100$ mm. For the double-slit experiment, $L_1=415$ mm and $L_2=385$ mm.}
\label{Fig.1}
\end{figure*}
As evident from the above discussion, the OAM of twisted photons can be argued to originate from the spiral motion of electrons \cite{Katoh_PRL2017}.
However, a theory suggesting that twisted photons can be generated from the rectilinear motion of electrons incident on a solid target, specifically through transition radiation, has been proposed \cite{Bogdanov_PRA2019,Bogdanov_JINST2020}.
Transition radiation is electromagnetic radiation produced when high-energy particles pass through an interface between materials with different permittivities \cite{Ginzburg_PhysUsp1996}. It is known to be emitted in both the forward and backward directions.
The study of transition radiation has a long history \cite{ginzburg1990transition} but so far the OAM properties of this radiation have not been investigated experimentally.
In the present study, we consider electrons being injected into a solid target tilted at 45$^{\circ}$ with respect to the beam direction, with the backward transition radiation emitted in the 90$^{\circ}$ direction to be detected.
According to the theory of Bogdanov \textit{et al}. \cite{Bogdanov_PRA2019,Bogdanov_JINST2020}, if the beam size on the solid target is sufficiently small, the produced transition radiation is expected to carry OAM.

In this Letter, we present a proof-of-principle experiment to test the theory of transition radiation carrying OAM \cite{Bogdanov_PRA2019,Bogdanov_JINST2020}.
Transition radiation was generated from the interaction of a 220 MeV electron beam with an Au-coated Si wafer.
To determine the OAM of the transition radiation, we recorded its diffraction patterns as it passed through a triangular aperture and a double slit.

According to the theory of twisted transition radiation \cite{Bogdanov_PRA2019,Bogdanov_JINST2020}, the probability of radiation of a twisted photon by a charged particle can be expressed as
\begin{equation}
\begin{aligned}
&dP_1(s,m,k_\perp,k_3)=e^2  \Big|\int^{\infty}_{-\infty}dt e^{-ik_0 t}\\
&\left(\dot{\spx}(t),\bs\Phi_{tw}(s,m,k_\perp,k_3;\spx(t))\right) \Big|^2 \left(\frac{k_\perp}{2k_0}\right)^3\frac{dk_3dk_\perp}{2\pi^2},
\end{aligned}
\end{equation}
where $\spx(t)$ represents the electron trajectory parameterized by the laboratory time $t$, $k_0$ is the photon energy, $k_3$ is the projection of the photon momentum on the direction of propagation, $k_\perp = \sqrt{k_0^2 - k_3^2}$, $m$ is the projection of the total angular momentum of the photon along the same axis, and $\bs\Phi_{tw}(s,m,k_\perp,k_3;\spx)$ is the mode function for the twisted photon without the normalization factor.
Here, we adopt a system of units in which $\hbar = 1$, the speed of light is $c = 1$, and $e^2 = 4\pi \alpha$, where $e$ is the electron charge and $\alpha$ is the fine structure constant.
When the condition $n_\perp = k_\perp / k_0 \ll 1$ is satisfied, the paraxial approximation applies.
In this approximation, the total angular momentum $m$ can be expressed as $m = l + s$.

In the study of Bogdanov \textit{et al}. \cite{Bogdanov_PRA2019}, $m$ is predicted to be sharply populated at certain values if the electron beam size on the target is sufficiently small, as follows:
\begin{equation}
k_\perp\sigma_\perp\ll1,
\end{equation}
where $\sigma_\perp$ is the transverse beam size on the target.
The theory shows that, for a helically microbunched beam, $m$ can take values of $|m| \geq 1$.
However, for a Gaussian beam of electrons, the total angular momentum of radiated photons satisfies $m = l + s = 0$, indicating that $l = -s$.
In this case, the transition radiation with $s = 1$ ($s = -1$) is predicted to possess $l = -1$ ($l = 1$).

Figure~1 shows a schematic of the experimental setup.
The experiment was conducted using a 220 MeV electron beam from the linear accelerator at the SAGA Light Source (SAGA-LS) synchrotron radiation facility in Japan.
This linear accelerator typically serves as an injector for the storage ring of the SAGA-LS and consists of six accelerating tubes.
The repetition rate of the beam acceleration was 1 Hz, and the average beam current was approximately 7.7 nA, with 34 microbunches.
Thus, the electron number per microbunch was $1.4 \times 10^9$.
In the present study, the phase of the radiofrequency (RF) signal for the first accelerating tube was optimized to compress the bunch length using a technique known as velocity bunching \cite{Serafini_AIP2001}.
The bunch length was estimated to be $\s_\mathrm{l} \approx 0.6$ mm.
The intensity of the detected transition radiation was proportional to the square of the beam current, suggesting that the transition radiation is coherent.

\begin{figure}[t]
\includegraphics[width=\hsize]{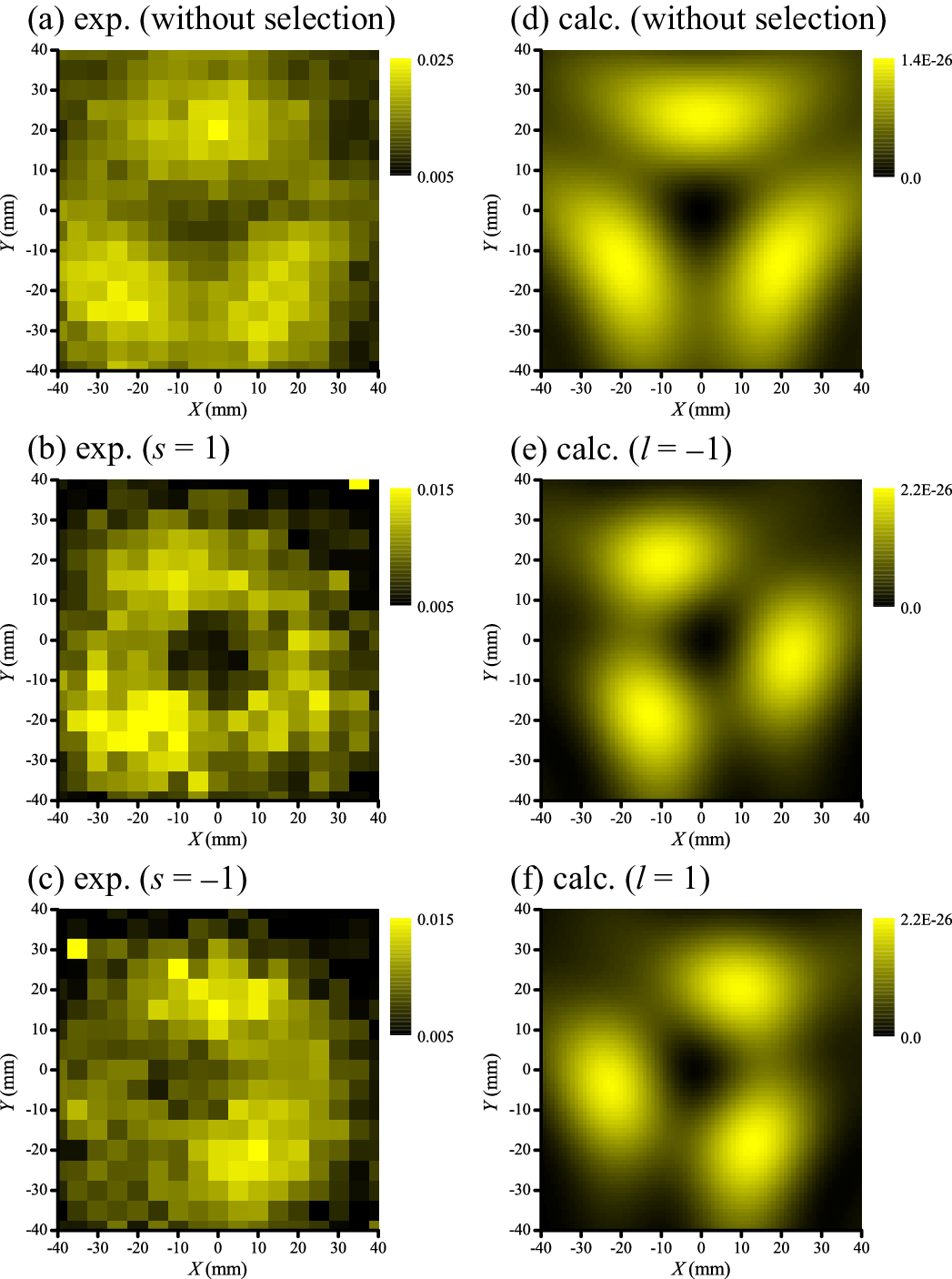}
\caption{Observed diffraction patterns for transition radiation passing through the triangular aperture for the cases (a) without SAM selection, (b) $s = 1$, and (c) $s = -1$. Calculated diffraction patterns for the cases (d) without OAM selection, (e) $l = -1$, and (f) $l = 1$.}
\end{figure}
An Au-coated Si wafer was used as the target in a vacuum chamber.
The diameter and thickness of the Si wafer were 4 inches (102 mm) and 525 $\mu$m, respectively.
The thickness of the Au coating was 100 nm.
The Si wafer was tilted 45$^{\circ}$ with respect to the beam direction, and the backward transition radiation emitted at 90$^{\circ}$ was extracted into the air through a quartz window.
The wafer was mounted onto a two-axis goniometer, which enabled precise alignment of the optical axis with respect to the triangular aperture and double slit.
The frequency (wavelength) of the transition radiation was selected using a bandpass filter.
Its central frequency was $f = 0.3$ THz (wavelength: $\lambda = 1$ mm), and the full-width at half-maximum was $\Delta f / f \approx 1/3$.
A combination of a quarter-wave plate and a linear polarizer was used to select the circular polarization, i.e., the spin angular momentum, of the transition radiation.
The quarter-wave plate was made from a crystal quartz plate designed for $\lambda = 1$ mm.
The linear polarizer was of the wire-grid type.
In experiments on twisted photons, the diffraction patterns for photons passing through various apertures, such as a single aperture, double slit, and triangular aperture, have been measured to determine the orbital angular momentum of twisted photons \cite{Hickmann_PRL2010,Goto_IR2019}.
In the present study, a triangular aperture and double slit were used.
A pyroelectric detector was used to detect sub-terahertz radiation.
The diameter of the sensitive area of the detector was 5 mm, and its sensitivity was 70 kV/W. A 60 $\mu$m-thick black polypropylene film was placed in front of the detector to eliminate infrared radiation.
The detector was mounted on a two-dimensional stage, and diffraction patterns were acquired by scanning the detector position in two dimensions.
The signal from the pyroelectric detector was digitized and acquired in coincidence with beam injection.
The horizontal and vertical beam sizes on the target were $\sigma_\mathrm{h} \approx 1.1$ mm and $\sigma_\mathrm{v} \approx 0.4$ mm, respectively.
Under the experimental conditions, $k_\perp \sigma_\mathrm{h} \approx 0.03$ and $k_\perp \sigma_\mathrm{v} \approx 0.01$, satisfying the condition of Eq.~(2).

\begin{figure}[t]
\includegraphics[width=\hsize]{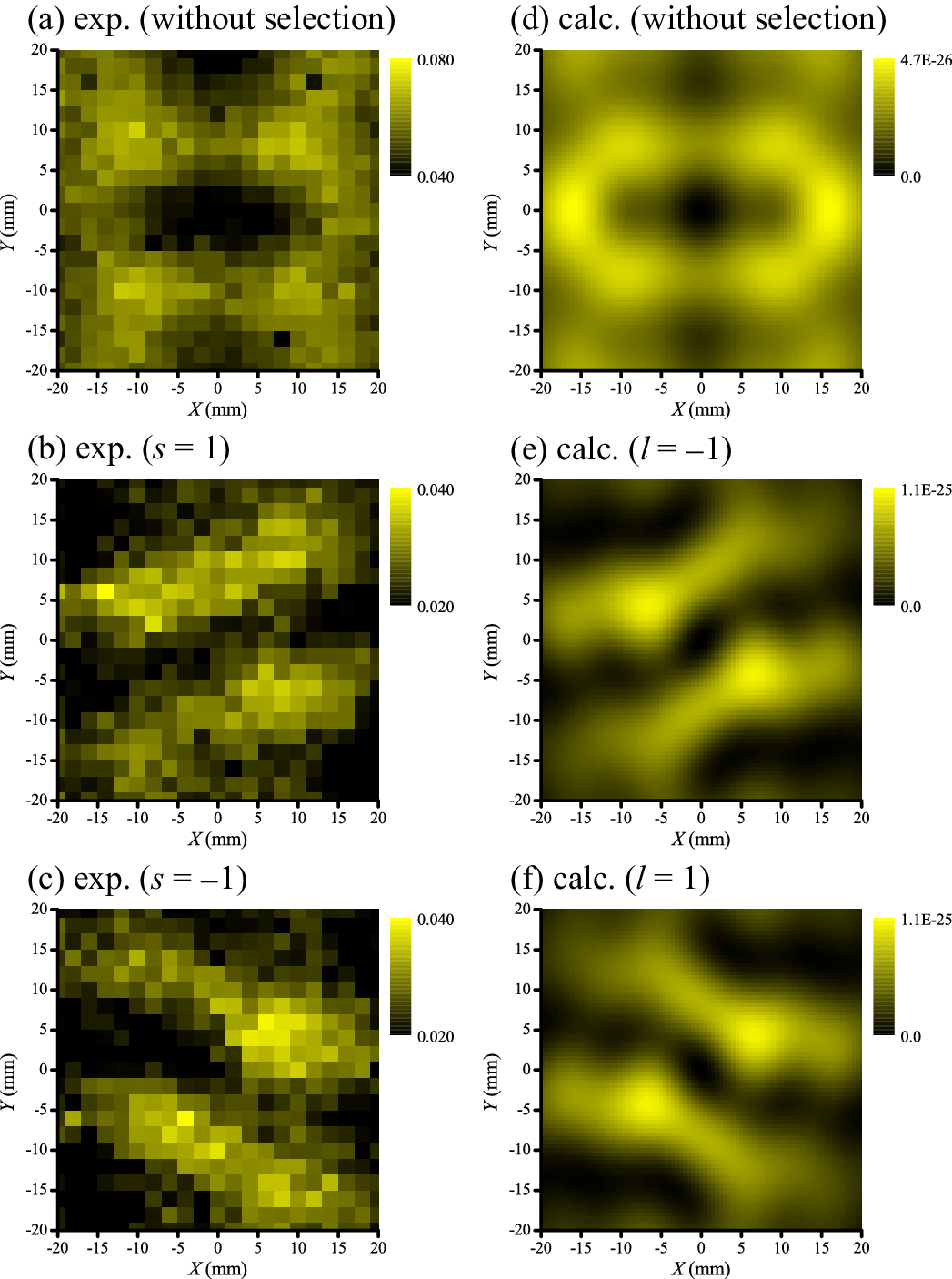}
\caption{Observed diffraction patterns for transition radiation passing through the double slit in the cases (a) without SAM selection, (b) $s = 1$, and (c) $s = -1$. Calculated diffraction patterns in the cases (d) without OAM selection, (e) $l = -1$, and (f) $l = 1$.}
\end{figure}
In Fig.~2, the acquired diffraction patterns for the transition radiation passing through the triangular aperture are shown for the cases (a) without SAM selection, (b) $s = 1$, and (c) $s = -1$.
The side length of the triangular aperture was 50 mm.
The distance from the radiation source to the triangle aperture was $L_1 = 500$ mm, and the distance from the triangle aperture to the detector was $L_2 = 1100$ mm.
The experimental data in Fig.~2(a) were acquired without the quarter-wave plate and linear polarizer shown in Fig.~1 (i.e., the SAM was not selected in these data).

Figures~2(d)--(f) show our calculated diffraction patterns for the transition radiation passing through the triangular aperture. The calculations were performed using the Fresnel approximation for different conditions. The calculated diffraction pattern in Fig.~2(d) closely reproduces the observed pattern in Fig.~2(a), validating the Fresnel approximation employed.
In Figs.~2(b) and 2(c), three predominant diffraction spots are observed, suggesting that the OAM carried by the transition radiation is $|l| = 1$ \cite{Hickmann_PRL2010,Goto_IR2019}.
In addition, the diffraction patterns acquired for $s = \pm 1$ were oppositely rotated with respect to the optical axis, in reasonable agreement with those calculated for $l = \mp 1$, respectively.
This result indicates that the transition radiation for $s = \pm 1$ carries OAM of $l = \mp 1$ and that the total angular momentum $m = l + s = 0$, as predicted by the theory of Bogdanov \textit{et al}. \cite{Bogdanov_PRA2019,Bogdanov_JINST2020}.

To confirm the obtained results, we conducted further experiments using a double slit.
Each slit had a square shape with dimensions of 10 mm $\times$ 40 mm, and the distance between the two slits was 20 mm.
The distance from the radiation source to the double slit was $L_1 = 415$ mm, and the distance from the double slit to the detector was $L_2 = 385$ mm.
Figures~3(a)--(c) show the measured diffraction patterns for the transition radiation passing through the double slit.
The diffraction pattern in Fig.~3(a) (without SAM selection) was symmetric with respect to the horizontal line.
However, the diffraction patterns for $s = \pm 1$ were tilted in opposite directions, suggesting that the OAM of the transition radiation is not zero.

The calculated diffraction patterns for the transition radiation are also shown in Fig.~3.
The diffraction patterns for the transition radiation observed for $s = \pm 1$ closely matched those calculated for $l = \mp 1$.
The prediction of the theory of Bogdanov \textit{et al}. \cite{Bogdanov_PRA2019, Bogdanov_JINST2020} was thus confirmed again.

In conclusion, we investigated the OAM of transition radiation produced by a 220 MeV electron beam incident on an Au-coated Si wafer.
The measurements were conducted under experimental conditions that satisfied the condition of Eq.~(2).
We determined the OAM of the transition radiation by acquiring the diffraction patterns for the transition radiation passing through a triangular aperture and a double slit.
From the observed diffraction patterns, we demonstrated that the transition radiation for $s = \pm 1$ carries $l = \mp 1$, respectively.
In the present work, the total angular momentum of the produced twisted light was $m = 0$.
From an application perspective, twisted photons with $|m| \ge 1$ are also valuable.
Bogdanov \textit{et al}. \cite{Bogdanov_PRA2019} proposed two primary methods for producing such twisted transition radiation.
One method involves using a helically microbunched beam as the incident beam, whereas the other uses a target with a spiral structure, such as cholesteric liquid crystals \cite{Bogdanov_PRE2021}.
Such studies would be of interest for future work.
According to the theory of Bogdanov \textit{et al}. \cite{Bogdanov_PRA2019,Bogdanov_ANNPHY2019}, Cherenkov radiation and edge radiation are also predicted to carry OAM.
The OAM of Cherenkov radiation is discussed in detail elsewhere \cite{Kaminer_PRX2016,ivanov2016quantum,Bogdanov_PRA2019}.
The present results encourage further exploration of twisted Cherenkov radiation and twisted edge radiation as the next step.

Katoh \textit{et al}. \cite{Katoh_PRL2017} claim that photons with OAM are ubiquitous in nature because they are produced by electrons in circular (spiral) motion.
From the present study, we conclude that twisted photons are even more ubiquitous because they are also produced by electrons moving in rectilinear motion.
For example, when cosmic rays such as muons or $\alpha$/$\beta$ particles emitted from radioisotopes strike substances on Earth, transition radiation is generated, and components of twisted photons are included in this radiation.

This work was supported in part by JSPS KAKENHI Grant Numbers JP21K12531 and JP24K15611.
The theoretical part of this study was supported by the Russian Science Foundation, grant No. 25-21-00283, https://rscf.ru/en/project/25-21-00283/.


%

\end{document}